\documentclass[submitting]{nst}

\usepackage{subfigure,dcolumn}

\usepackage[T2A,T1]{fontenc}
\usepackage[russian,english]{babel}

\usepackage{listings}
\usepackage{multirow}
\usepackage{makecell}
\begin{document}

\title{Reducing Systematic Bias in Machine Learning Applications to Signal Extraction in High-Energy Nuclear Physics}\thanks{Supported by National Key R\&D Program of China under grant No. 2024YFA1610803 and the National Natural Science Foundation of China under grant Nos. 12361141827 and 12175223.}

\author{Yan Wang}
\affiliation{Department of Modern Physics, University of Science and Technology of China, Hefei 230026, China}

\author{Rangrong Ma}
\affiliation{Physics Department, Brookhaven National Laboratory, Upton, New York 11973, USA}

\author{Kaifeng Shen}
\affiliation{Department of Modern Physics, University of Science and Technology of China, Hefei 230026, China}

\author{Zebo Tang} \email[Corresponding author: ]{zbtang@ustc.edu.cn}
\affiliation{Department of Modern Physics, University of Science and Technology of China, Hefei 230026, China}

\author{Wangmei Zha} \email[Corresponding author: ]{first@ustc.edu.cn}
\affiliation{Department of Modern Physics, University of Science and Technology of China, Hefei 230026, China}

\begin{abstract}
Machine learning techniques are increasingly being applied in high-energy nuclear physics data analysis thanks to their outstanding performance. One key challenge in such applications is the construction of training samples that can accurately represent real data. Training samples are typically generated through detector simulations, but discrepancies between simulated and real data can lead to degradation in machine learning performance and systematic biases in the results. This paper introduces two methods: i) cumulative distribution function mapping and ii) shift-and-scale, to align simulated signals with real data, which can aid in eliminating the aforementioned issues. We use the J/$\psi$ yield measurement in 200 GeV Ru+Ru and Zr+Zr collisions with the STAR experiment as an example to demonstrate the application and effectiveness of the proposed methods.
\end{abstract}


\maketitle

\section{Introduction}
 
Machine learning has become an indispensable tool across a wide range of scientific and technological fields, from image recognition and natural language processing to recommendation systems and speech analysis. Its strength lies in the ability to identify complex patterns in large datasets and to provide powerful classification, regression, and clustering capabilities~\cite{LeCun:2015pmr, PANG:2020lda, Jamal:2025gjy, Li:2025csc, He:2023urp, Li:2025gnw, Niu:2019pro}. In the field of nuclear physics, machine learning offers significant advantages for extracting key physical observables such as nuclear charge radii, nuclear masses, and nuclear binding energies~\cite{BoZhou:2023, Qing-FengLi:2021, LuTang:2024, Zhong-ZhouRen:2024, Guo-QiangZhang:2024}. Moreover, in high-energy nuclear physics involving relativistic heavy-ion collisions, machine learning is particularly valuable in tasks that involve pattern classification, such as distinguishing physics signals from background~\cite{Albertsson:2018maf, CMS:2020poo, ALICE:2024oob, ALICE:2022cop, ALICE:2022exq, ALICE:2023wbx}. Commonly used classification algorithms include boosted decision trees (BDT)~\cite{Coadou:2022nsh, Roe:2004na, Lalchand:2020dej}, deep neural networks~\cite{Schmidhuber:2014bpo, Madrazo:2017qgh, Baldi:2016fql}, and ensemble methods~\cite{Ensemble, Bentley:2024bnm}, which can exploit multi-dimensional correlations among observables to enhance signal significance beyond what is achievable with straight cut based approaches. Regardless of the specific algorithm employed, the training of a reliable classifier requires well-defined signal and background samples. While background distributions can often be obtained directly from data, signal samples are typically generated through Monte Carlo simulations. However, such simulations are inherently imperfect and may not accurately reproduce the feature distributions observed in real data. Training on such mismatched simulated samples can mislead the classifier, degrade its performance, and introduce systematic biases, for example in signal yield measurements.

To address this challenge, we propose two methods: cumulative distribution function (CDF) mapping and shift-and-scale. The former systematically aligns simulated sample with real data without requiring explicit parameterization of the shapes of feature distributions, providing a robust and flexible way to eliminate mismatches between simulated and real data. The latter works well if the feature distributions reasonably resemble those in real data. Compared to the CDF method, the shift-and-scale procedure can be applied within a smaller scope, but easier to implement. 

J/$\psi$ is an important probe for investigating the properties of Quark–Gluon Plasma~\cite{Matsui:1986dk, Gyulassy:2004vg}. The measurement of its yield has attracted widespread attention~\cite{STAR:2013eve, STAR:2012wnc, STAR:2016utm, PHENIX:2006gsi, ALICE:2013osk, ALICE:2015nvt, STAR:2019fge, Wang:2024pjc, Tang:2020ame}. In this paper, we utilize BDT implemented through the eXtreme Gradient Boosting (XGBoost) framework~\cite{chen2016xgboost} to separate J/$\psi$ signal from background in Ru+Ru and Zr+Zr collisions at the center-of-mass energy per nucleon-nucleon pair ($\sqrt{s_{\rm NN}}$) of 200 GeV recorded by the STAR experiment at RHIC. This case study illustrates both the implementation of the proposed correction methods and their effectiveness in improving the robustness of machine learning–based signal extraction. Details about J/$\psi$ reconstruction are introduced in Sec. \ref{sect::jpsireco}, model construction, training and application are presented in Secs. \ref{sect::trainingsample} and \ref{sect::modeltraining}, model validation through a series of self-consistency checks and results are discussed in Sec. \ref{sect::modelvalidation}, and a summary is provided in Sec. \ref{sect::summary}.

\section{J/$\psi$ Reconstruction}
\label{sect::jpsireco}

In Ru+Ru and Zr+Zr collisions, J/$\psi$ mesons are reconstructed via the dielectron decay channel~\cite{STAR:2012wnc, Tang:2020ame, Chen:2024aom}, with electron \footnote{Denote both electron and positron unless specified otherwise} identification based on the Time-Of-Flight (TOF)~\cite{Llope:2009zz}, the Time Projection Chamber (TPC)~\cite{STAR:1997sav,Anderson:2003ur}, and the Barrel Electromagnetic Calorimeter (BEMC)~\cite{STAR:2002ymp} detectors. Loose preselection cuts are usually applied on measured particle identification (PID) variables to preliminarily suppress the background while retaining a sufficient amount of signal, enabling subsequent application of machine learning techniques to further enhance the signal significance. 

\begin{figure}[htbp]
\centering
\includegraphics[width=0.46\textwidth,clip]{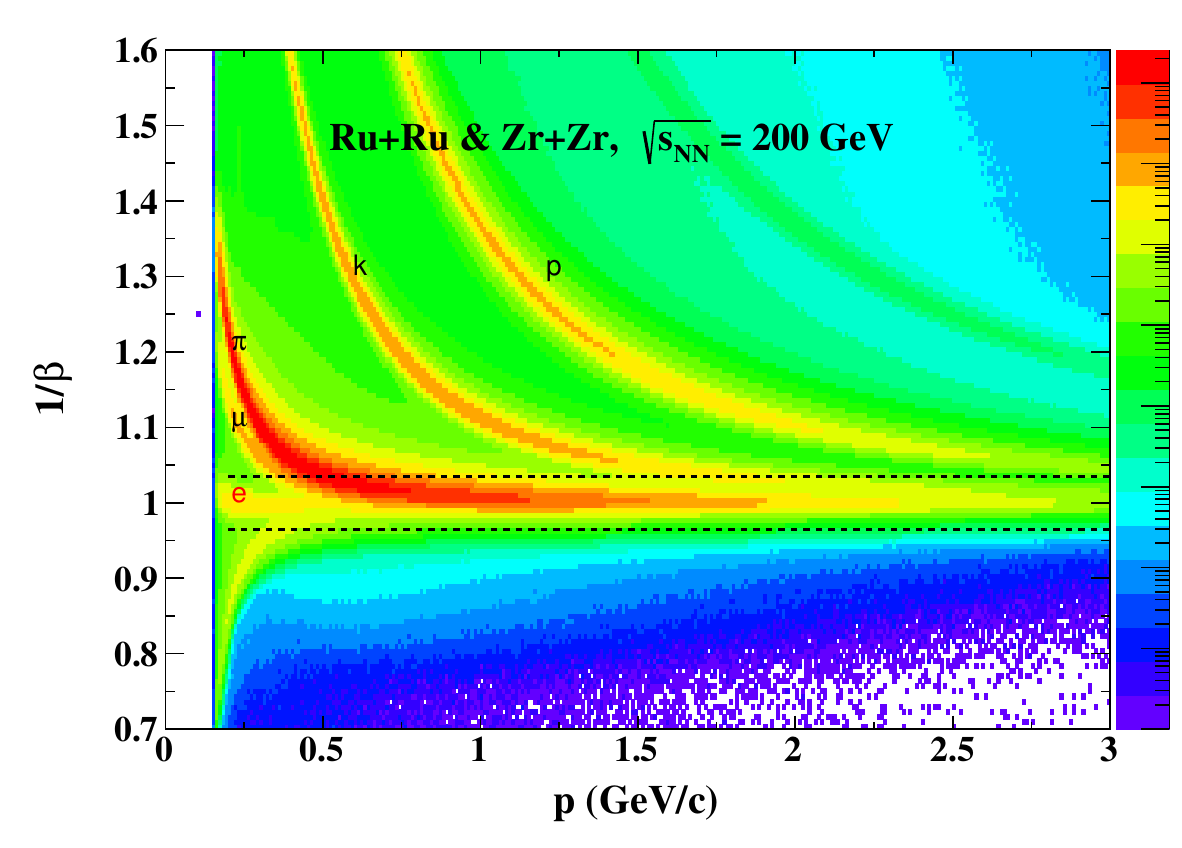}
\caption{Distribution of $1/\beta$ versus momentum for charged particles, where the dashed horizontal lines indicate the $1/\beta$ cuts for electron preselection.}
\label{fig_beta}      
\end{figure}

TOF measures the arrival time of a charged particle. Combined with the collision start time and charged particle momentum measured in the TPC, the variable $1/\beta= c/v$ can be calculated, where $c$ is the speed of light and $v$ is the speed of the charged particle. The distribution of $1/\beta$ versus momentum ($p$) for charged particles is shown in Fig.~\ref{fig_beta}. Particles with different masses possess different velocities at a given momentum, resulting in distinct bands as seen in the figure which can be used to perform PID. A preselection cut of $\left|1/\beta - 1 \right|<0.035$ is applied, shown as dashed lines in Fig.~\ref{fig_beta} and listed in Tab.~\ref{table_pre_pid}. 


\begin{figure}[htbp]
\centering
\includegraphics[width=0.46\textwidth,clip]{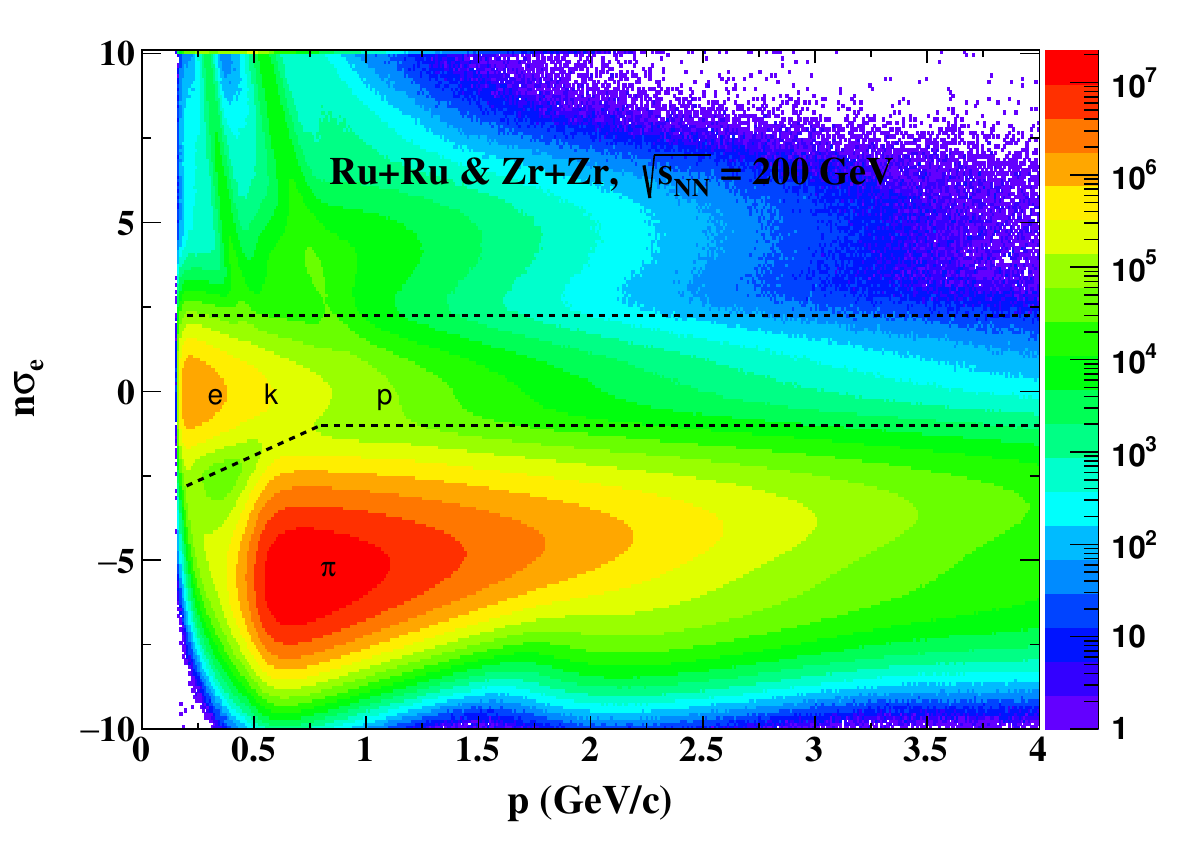}
\caption{Distribution of $n\sigma_{e}$ versus momentum for charged particles after applying the $\left|1/\beta - 1 \right|<0.025$ cut. The dashed curves correspond to the $n\sigma_{e}$ cuts used for electron preselection.}
\label{fig_nsigame}      
\end{figure}

The TPC measures the ionization energy loss per unit length ($dE/dx$), which can be used to distinguish electrons from other hadrons. Particle identification commonly utilizes the normalized deviation of the measured $dE/dx$ from its theoretical expectation for different particle species, denoted as $n\sigma_{e}$, $n\sigma_{\pi}$, and so on. For example, for electrons, the $n\sigma_{e}$ distribution is expected to follow a Gaussian distribution of mean equal to 0 and width equal to 1. Figure~\ref{fig_nsigame} shows $n\sigma_{e}$ versus $p$ distribution for charged particles after applying the TOF cut of $\left|1/\beta - 1 \right|<0.025$. The dashed curves indicate the momentum-dependent preselection cuts applied in preparation for subsequent machine learning. 

\begin{figure}[htbp]
\centering
\includegraphics[width=0.46\textwidth,clip]{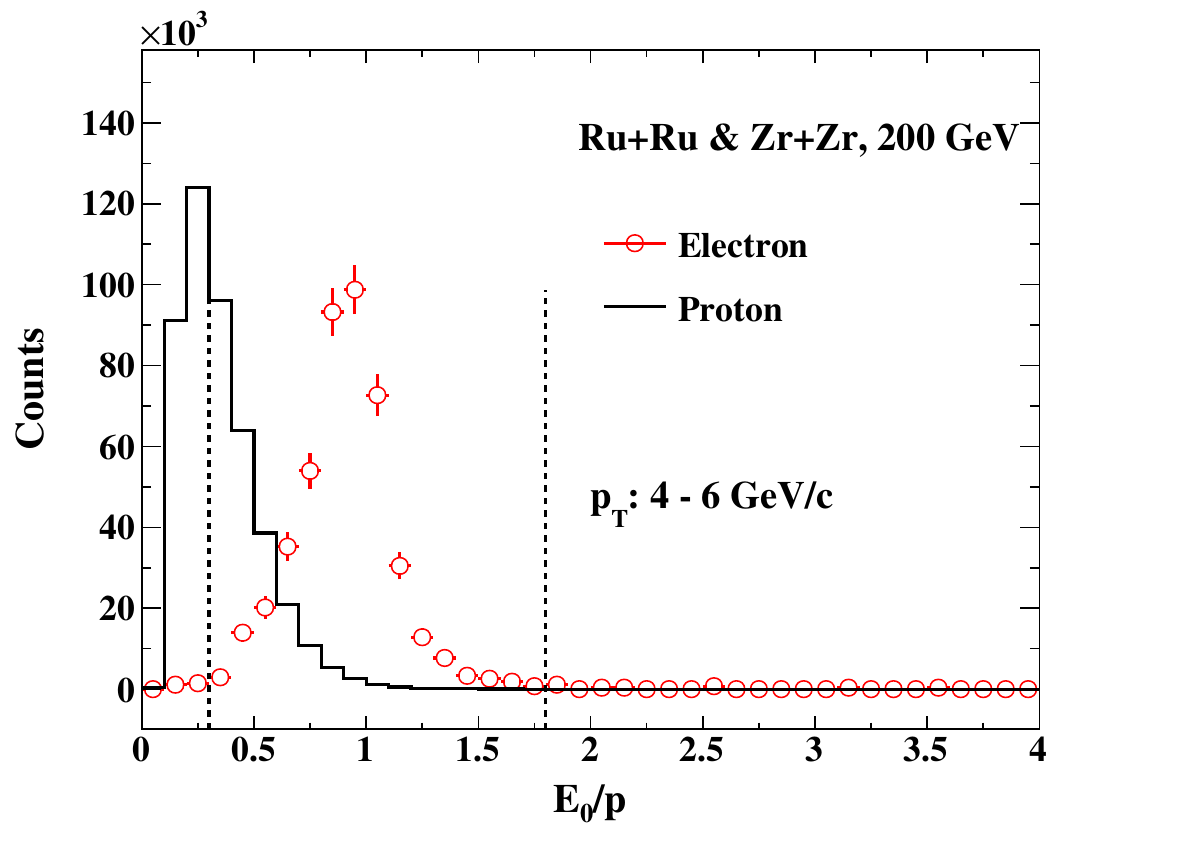}
\caption{Distributions of $E_{0}/p$ for electrons (circles) and protons (solid line) within $4<p_{\rm T}<6$ GeV/$c$. The vertical dashed lines correspond to the $E_{0}/p$ cuts used for electron preselection.}
\label{fig_E0Op}      
\end{figure}

The BEMC achieves PID at high $p$ through the ratio of the total energy deposited in the BEMC by an incident particle to its momentum ($E_{tot}/p$). However, electron identification commonly uses the largest tower energy within a BEMC cluster ($E_{0}$), as it offers better resolution and is less impacted by high-occupancy environments in heavy-ion collisions compared to using $E_{tot}$. Figure~\ref{fig_E0Op} illustrates the $E_{0}/p$ distributions for protons and electrons within the transverse momentum ($p_{\rm T}$) range of 4 to 6 GeV/$c$. Here, the electron sample, usually referred to as photonic electrons, is obtained from photon conversions and neutral meson Dalitz decays~\cite{STAR:2015tnn}. As expected, the electron distribution peaks at a larger $E_{0}/p$ value than that of the protons. The dashed vertical lines denote the preselection $E_{0}/p$ cuts.

\begin{figure*}[htbp]
    \centering
    \subfigure[With preselection cuts listed in Tab.~\ref{table_pre_pid}]{
        \includegraphics[width=0.32\textwidth]{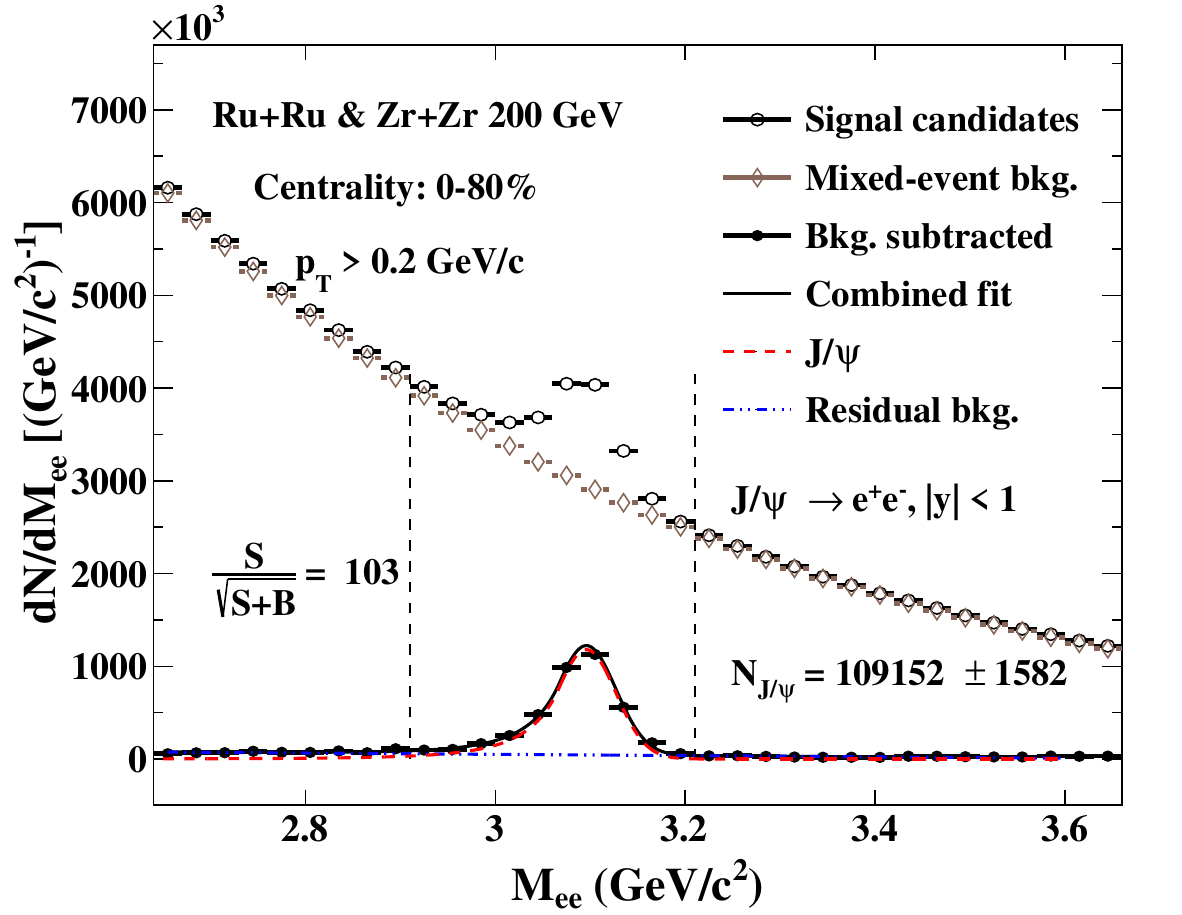}
    }
    \hfill
    \subfigure[With preselection and BDT cuts listed in Tab.~\ref{table_pre_pid}]{
        \includegraphics[width=0.3\textwidth, trim=33 0 0 0, clip]{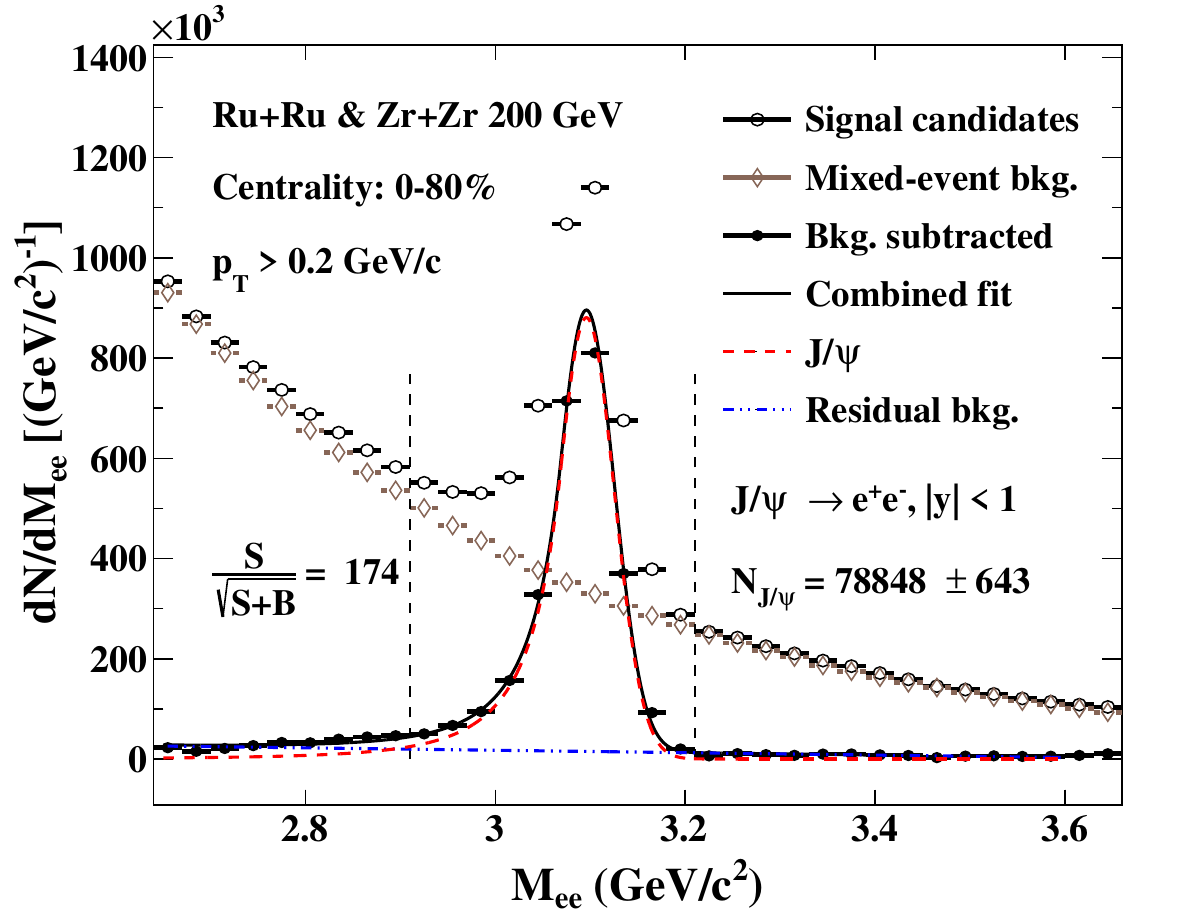}
    }
    \hfill
    \subfigure[With straight cuts listed in Tab.~\ref{table_pid}]{
        \includegraphics[width=0.3\textwidth, trim=33 0 0 0, clip]{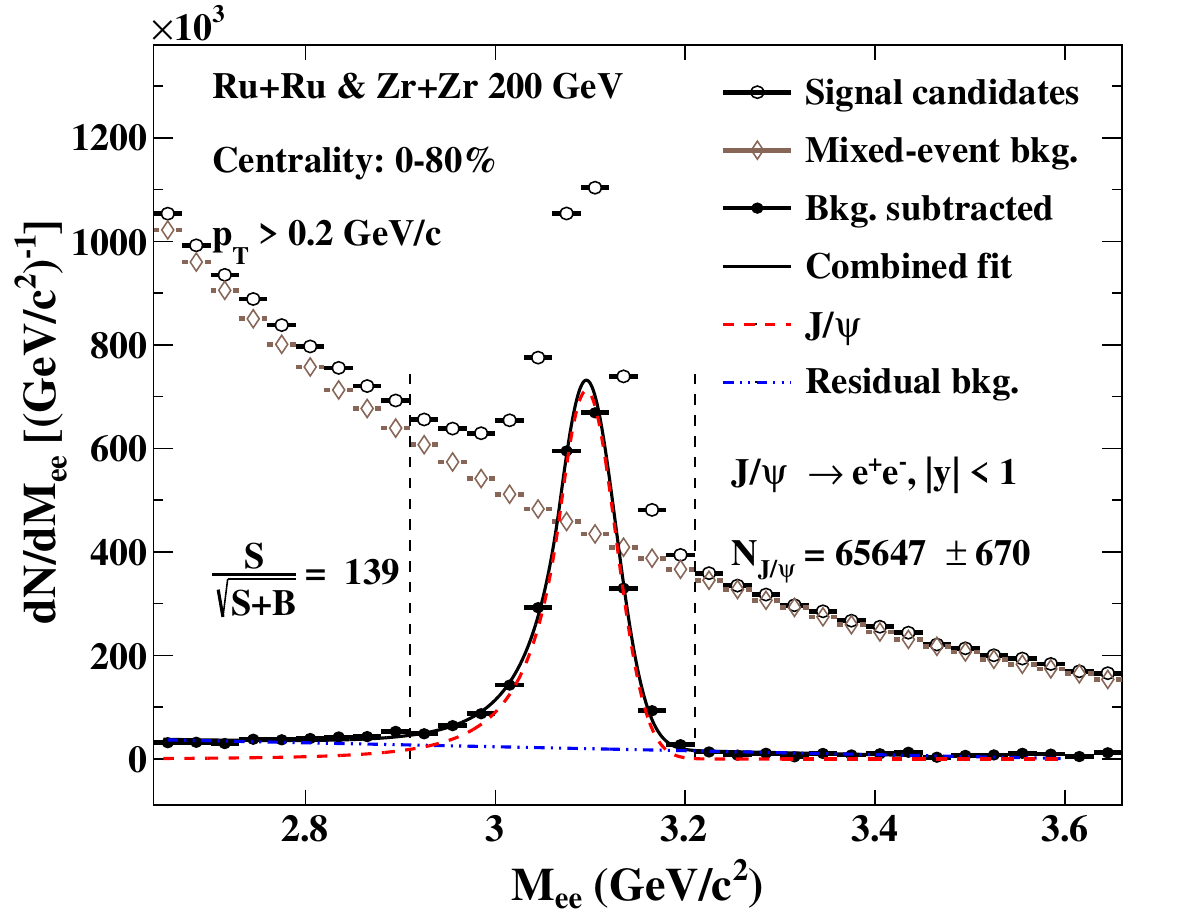}
    }

    \caption{Invariant mass distributions of J/$\psi$ candidates reconstructed via the dielectron channel, within $|y| < 1$ and $p_{\rm T} > 0.2$ GeV/$c$, for Ru+Ru and Zr+Zr collisions. Different panels correspond to different electron identification cuts.}
    \label{fig_invariant}
\end{figure*}

The application of aforementioned TOF, TPC and BEMC preselection PID cuts depends on $p_{\rm T}$ and $p$ ranges of the charged particle and whether it leaves signals in TOF and BEMC, as summarized in Tab.~\ref{table_pre_pid}. Electrons and positrons, identified based on these preselection cuts, can be paired to form J/$\psi$ candidates, and the invariant mass distribution of these candidates is shown in panel (a) of Fig.~\ref{fig_invariant}. The combinatorial background is estimated via the mixed-event technique~\cite{PHENIX:2009gyd}. After subtracting the combinatorial background, the invariant mass distribution of J/$\psi$ candidates is fit with a function (black line) composed of two components: a Crystal Ball function (red dashed line) describing the J/$\psi$ signal, and a linear function (blue dashed line) accounting for the residual background from semi-leptonic decays of heavy-flavor quark pairs and Drell-Yan processes. The raw J/$\psi$ yield can be extracted by counting the number of electron-positron pairs within the mass range of [2.91, 3.21] GeV/$c^2$, as indicated by the vertical dashed lines in Fig.~\ref{fig_invariant} (a), and subtracting the contribution of the residual background obtained from the fit. A total of about 110k J/$\psi$ are obtained, and the significance of the J/$\psi$ signal is 103. Building upon this, we will apply machine learning techniques to further enhance the signal significance.

\begin{table*}[htbp]
    \centering
    \caption{Electron preselection cuts and the optimal BDT cut criterion from machine learning.}
    \begin{tabular*}{\textwidth}{@{\extracolsep{\fill}} cccc}
        \hline
        Track $p_{\rm T}$ & Detectors used & Electron preselection cuts & BDT criterion \\
        \hline
        $p_{\rm T} \leq 1.0$ GeV/$c$ & TPC, TOF & {\footnotesize\makecell{$\left|1/\beta - 1 \right|<0.035$; \\ for $p > 0.8$ GeV$/c$: $-1 < n\sigma_{e} < 2.25$; \\ for $p \leq 0.8$ GeV$/c$: $(3\times p - 3.4) < n\sigma_{e} < 2.25$; }} & BDT $>$ 0.7\\
        \hline
        \multirow{7}{*}{$p_{\mathrm{T}} > 1.0$ GeV/$c$} & \makecell{TPC, TOF and \\ BEMC no matching} & {\footnotesize\makecell{$\left|1/\beta - 1 \right|<0.035$; \\  $-1 < n\sigma_{e} < 2.25$ }} & \multirow{7}{*}{BDT $>$ 0.7}\\
          & \multirow{3}{*}{\makecell{TPC, BEMC and \\ TOF no matching}} & \multirow{3}{*}{\footnotesize\makecell{$-1.25 < n\sigma_{e} < 2.25$; \\ $0.3 < E_{0}/p < 1.8$}} \\
          & & \\
          & & \\
         & \makecell{TPC, TOF and \\ BEMC} & {\footnotesize\makecell{$\left|1/\beta - 1 \right|<0.035$; \\  $-1.75 < n\sigma_{e} < 2.25$; \\ $0.3 < E_{0}/p < 1.8$ }} \\
        \hline
    \end{tabular*}
    \label{table_pre_pid}
\end{table*}

\section{Construction of Training Samples}
\label{sect::trainingsample}

For J/$\psi$ reconstruction, the classification targets are J/$\psi$ candidates composed of electron-positron pairs. To determine whether a candidate corresponds to a signal or background, specific features are selected. In principle, any quantity that can aid in the unbiased discrimination between signal and background can potentially serve as a feature. In this analysis, the selected features include $n\sigma_{e}$, $n\sigma_{\pi}$, $E_{0}/p$, and DCA (Distance of Closest Approach, defined as the minimum distance between a charged particle’s trajectory and the collision vertex) for both the electron and positron, which are labeled as ``tr1" and ``tr2", respectively. 

\begin{figure}[htbp]
\centering
\includegraphics[width=0.46\textwidth,clip]{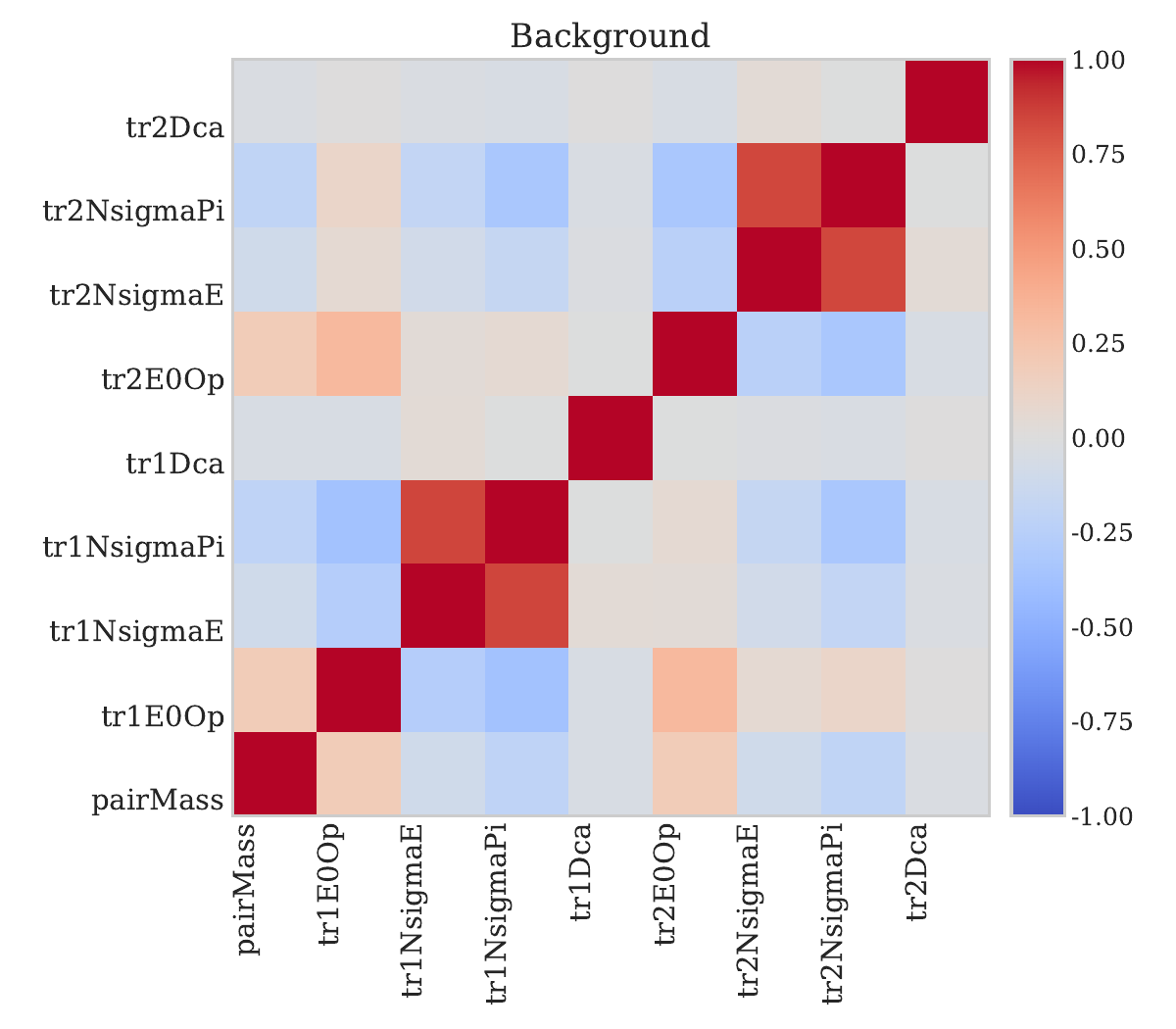}
\caption{Correlations between selected features and the electron-positron pair invariant mass for the background training sample.}
\label{fig_corr_bkg_check}      
\end{figure}

The background training sample is selected from electron-positron pairs within the invariant mass intervals of (2.65, 2.85), (3.25, 3.5), and (3.8, 4.5) GeV/$c^2$, away from the J/$\psi$ signal. The interval (3.5, 3.8) GeV/$c^2$ is excluded due to the presence of a small number of $\psi$(2S) mesons. This approach, referred to as the sideband method, uses background events from the sideband regions of the signal to approximate the background events under the J/$\psi$ peak. Consequently, it is required that the features used for classification do not strongly depend on the invariant mass. This serves as one of the key principles for feature selection. To verify this, the correlation between selected features and the invariant mass for the background sample is shown in Fig.~\ref{fig_corr_bkg_check}, with no significant correlations observed. Notably, $n\sigma_{e}$ and $n\sigma_{\pi}$ exhibit a strong correlation as both are derived from $dE/dx$.

\begin{figure*}[htbp]
    \centering
    \subfigure[$n\sigma_{e}$ distributions]{
        \includegraphics[width=0.45\textwidth]{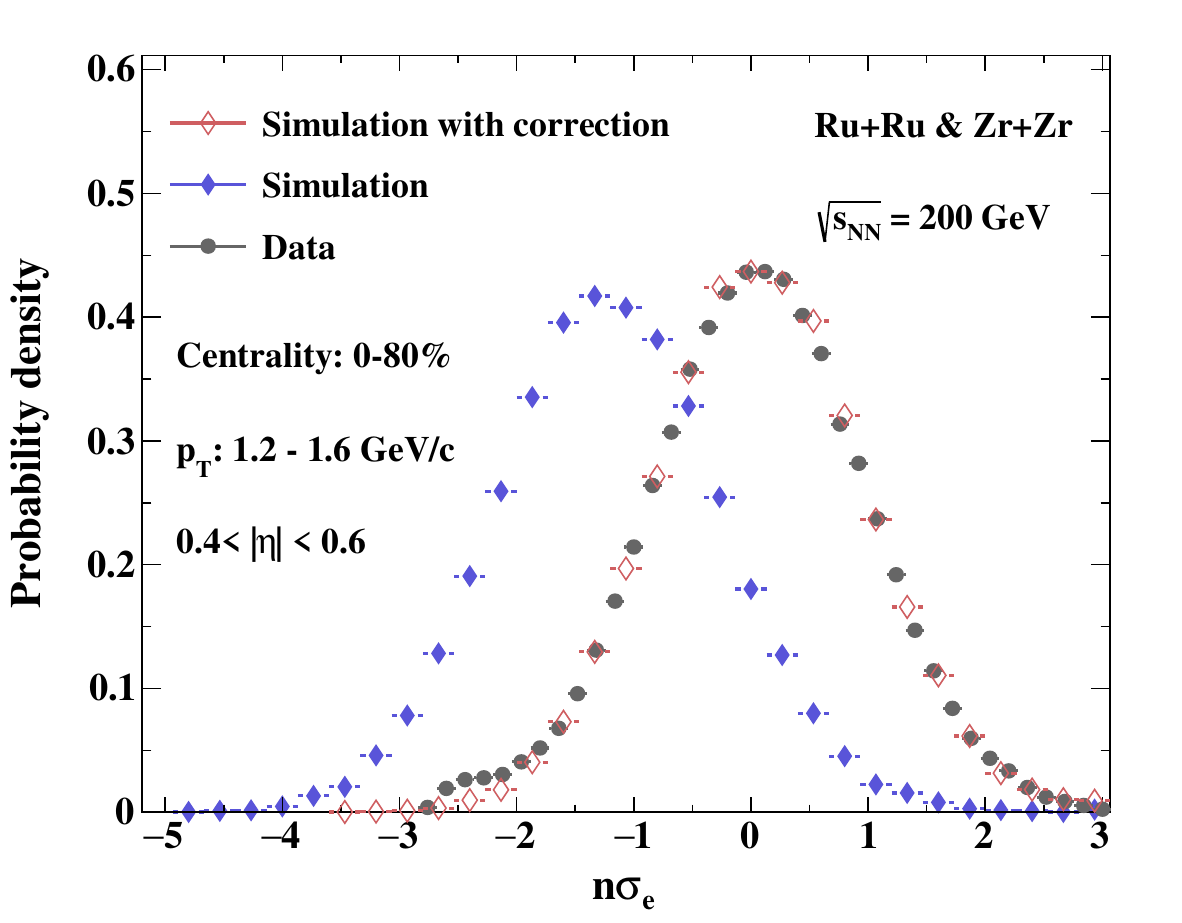}
    }
    \hfill
    \subfigure[$n\sigma_{\pi}$ distributions]{
        \includegraphics[width=0.45\textwidth]{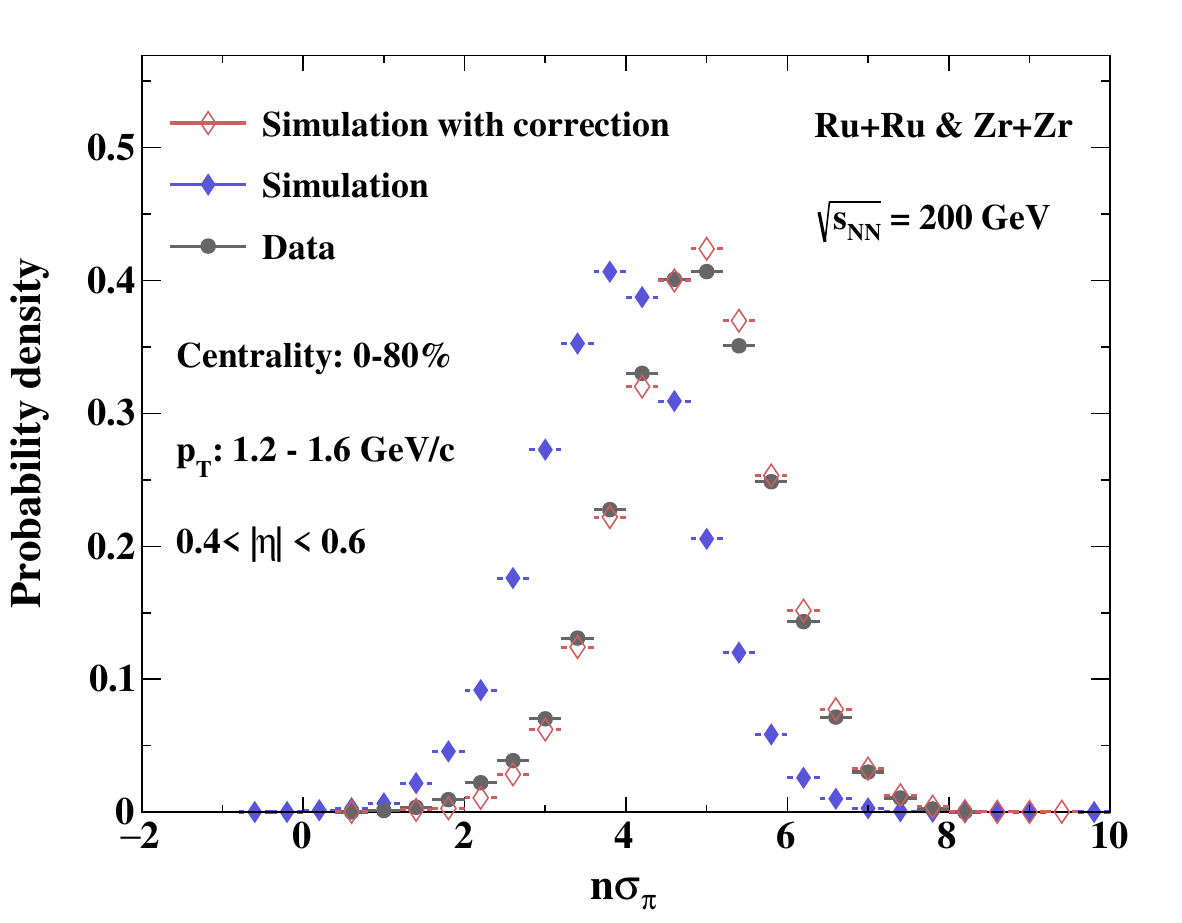}
    }

    \subfigure[$E_{0}/p$ distributions]{
        \includegraphics[width=0.45\textwidth]{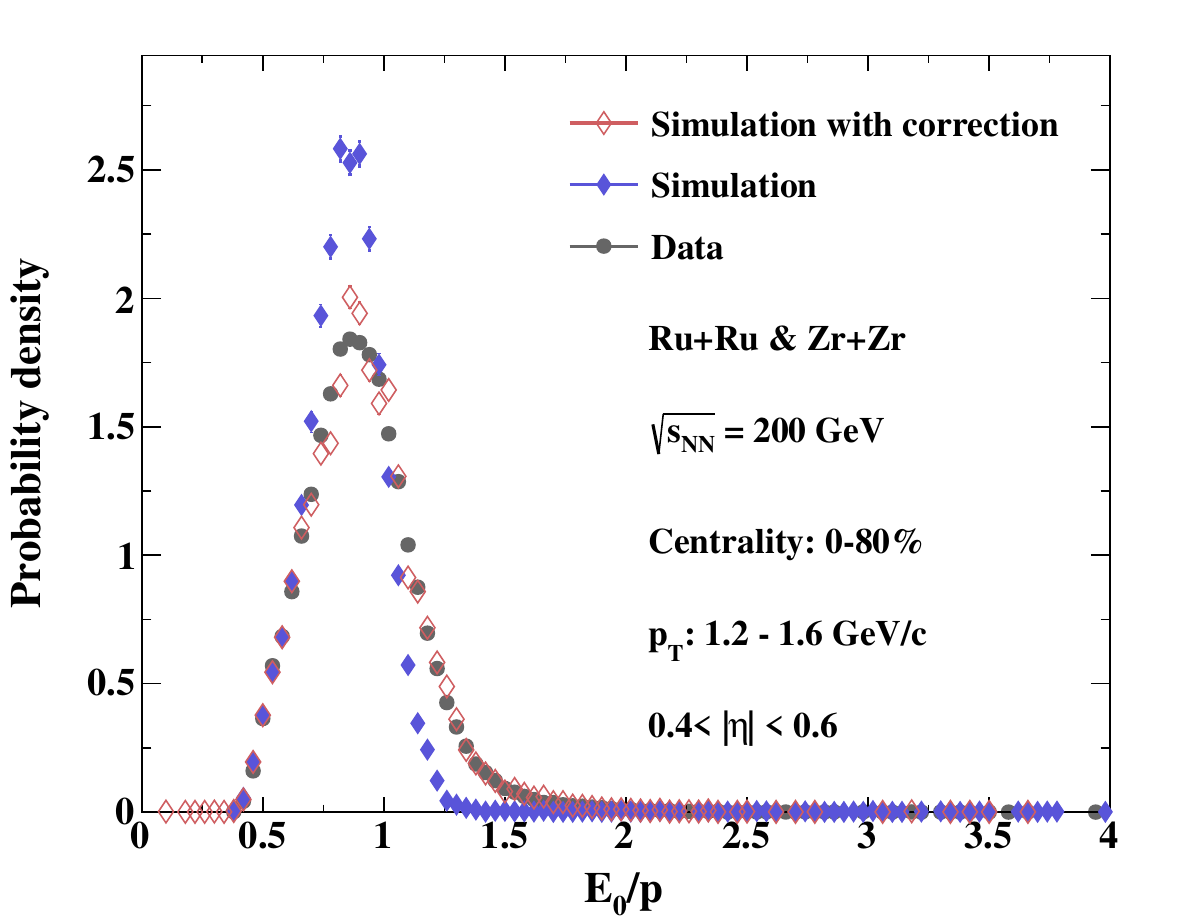}
    }
    \hfill
    \subfigure[DCA distributions]{
        \includegraphics[width=0.45\textwidth]{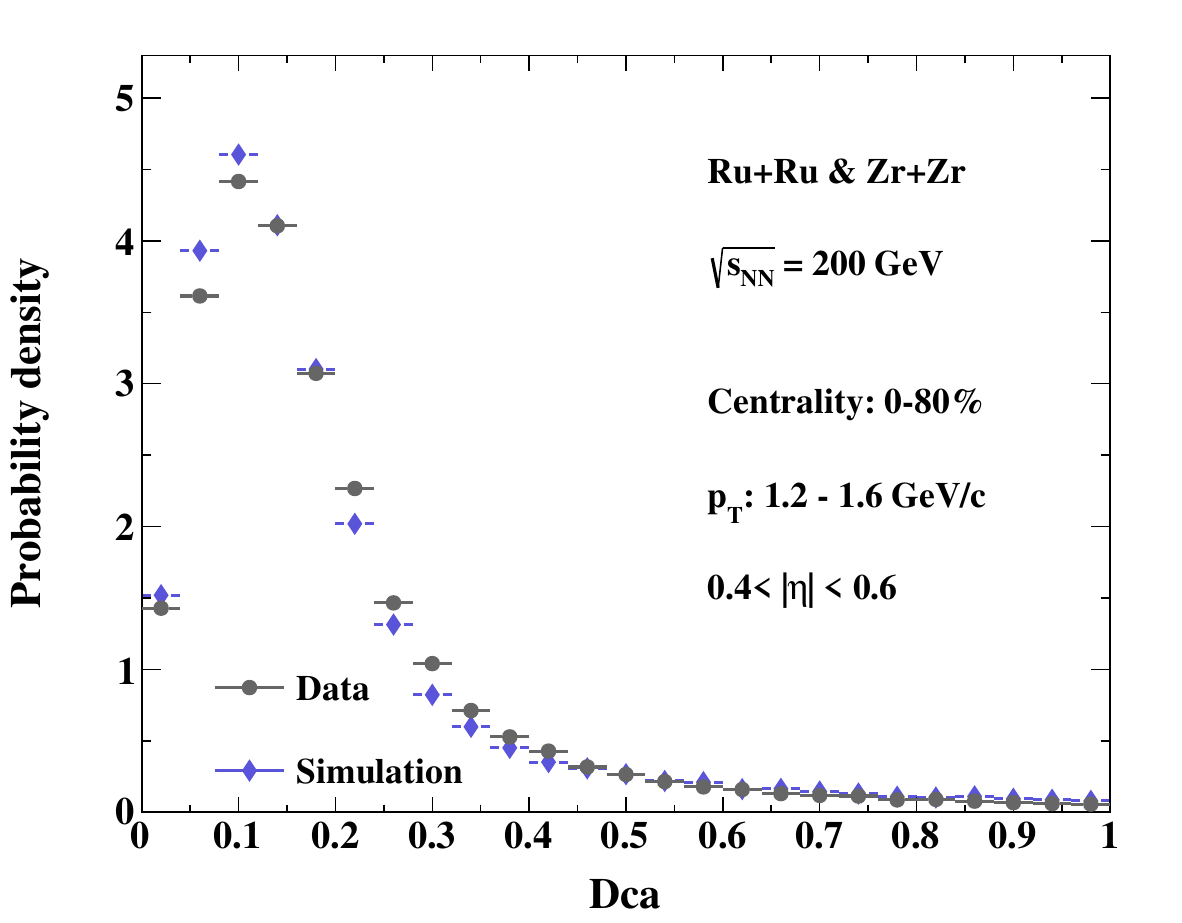}
    }

    \caption{Comparisons of feature distributions for electrons within $1.2<p_{\rm T} <1.6$ GeV/$c$ and $0.4<|\eta|<0.6$ from simulations, both with (open diamonds) and without (filled diamonds) corrections, to those from real data (filled circles).}
    \label{fig_feature}
\end{figure*}

The signal training sample is generated through Monte-Carlo routines, where electrons and positrons from J/$\psi$ decays are propagated through a GEANT3-based~\cite{STAR:2008med} simulation of the STAR detector. These simulated tracks are then embedded into real Ru+Ru and Zr+Zr collision events, which are subsequently reconstructed using the same procedure as in real data. It is perceivable that the simulation could not perfectly reproduce real data due to complexities in detector conditions varying over time and high-occupancy heavy-ion environment. The performance of the simulation can be examined by comparing the $n\sigma_{e}$, $n\sigma_{\pi}$, $E_{0}/p$ and DCA distributions for electrons from simulated J/$\psi$ decays with those of photonic electrons in real data, as illustrated in Fig.~\ref{fig_feature}. While the DCA distributions agree reasonably well between simulation and real data, noticeable shifts are observed in $n\sigma_{e}$ and $n\sigma_{\pi}$ distributions. For $E_{0}/p$ distributions, a clear difference in the shape is seen. These discrepancies can degrade the performance of machine learning and introduce potential biases in the measurement, and therefore need to be addressed before model training.

Since the shapes of the \(n\sigma_e\) and $n\sigma_{\pi}$ distributions for electrons in simulated data and real data are similar, a shift-and-scale method can be used. It is implemented as follows:
\begin{equation}
\setlength{\abovedisplayskip}{6pt}
\setlength{\belowdisplayskip}{6pt}
\begin{gathered}
    n\sigma_{i,cor} = (n\sigma_{i} - \mu_{simu}) \cdot (\sigma_{data}/\sigma_{simu}) + \mu_{data},
\end{gathered}
\label{func-re-calibration} 
\end{equation}
where $\mu_{data}$ ($\sigma_{data}$) and $\mu_{simu}$ ($\sigma_{simu}$) denote the mean values (standard deviations) of the $n\sigma_{i}$ distributions for electrons from data and simulation, respectively, and $i$ could be $e$ or $\pi$. After applying the shift-and-scale correction, the $n\sigma_{e}$ and $n\sigma_{\pi}$ distributions of the simulated electrons are brought into agreement with those of real data, as shown in Fig.~\ref{fig_feature} (a) and (b). 

For $E_{0}/p$ distributions where a difference in shape between real data and simulation is seen, the shift-and-scale method would not work. So, a cumulative distribution function (CDF) mapping method is developed. Let $f_{\text{simu}}(x)$ and $f_{\text{data}}(x)$ ($x \in [x_{\min}, x_{\max}]$) denote the probability density functions (PDFs) of feature $x$ in simulation and real data, respectively. The corresponding CDF is defined as:
\[
C(x) = \int_{x_{\min}}^{x} f(x') \, dx'.
\]
In the simulation, for feature $x$ with value $A$, we compute its cumulative probability as
\[
C_{\text{simu}}(A) = \int_{x_{\min}}^{A} f_{\text{simu}}(x') \, dx'.
\]
We then search for a value $B$ in the data distribution such that:
\[
C_{\text{data}}(B) = C_{\text{simu}}(A) \quad \Rightarrow \quad B = C_{\text{data}}^{-1}(C_{\text{simu}}(A)).
\]
This defines a mapping from the original value $A$ to a corrected value $B$, such that the transformed simulation distribution matches the data distribution. After applying the CDF mapping, the corrected $E_0 / p$ distribution in simulation (open diamonds) aligns well with that from real data (solid circles), as shown in Fig.~\ref{fig_feature} (c). Considering that PDFs in practical applications are discrete histograms, we implement this method numerically by computing discrete CDFs and searching for the closest cumulative value. Consequently, when the distribution from simulation differs significantly in shape from that in real data, fluctuations of the size of one bin can occur during the matching process, manifesting as an increase in statistics in a bin while adjacent bins experience statistics loss. This is the reason for the statistical anomalies observed in some bins of the $E_0 / p$ distribution after CDF mapping in Fig.~\ref{fig_feature} (c). Since this artifact is caused by histogram binning, it does not affect model training for machine learning which takes signal and background samples in continuous values. One can effectively reduce such fluctuations by decreasing the bin width.

Compared to the shift-and-scale method, the CDF mapping method offers greater generality in its application even though it comes with more complexities in implementation. Importantly, both methods can preserve the correlations among different features, a key aspect for machine learning techniques to outperform straight cuts, while making simulated sample closely align with real data. This is also the reason why signal samples cannot be generated by randomly sampling distributions of individual features from real data, which would have lost all the correlations.

\section{Model Training and Application}
\label{sect::modeltraining}

\begin{figure}[htbp]
\centering
\includegraphics[width=0.46\textwidth,clip]{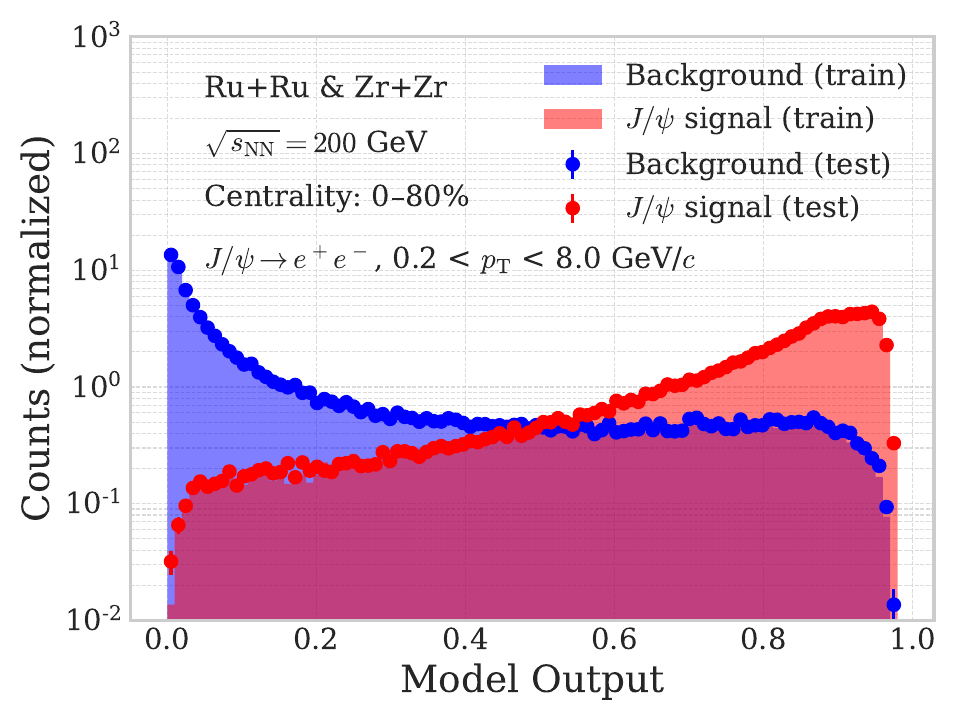}
\caption{Model output distributions for both the training and testing samples. The red (blue) histogram represents the training signal (background) sample, while the red (blue) points are for the testing signal (background) sample.}
\label{fig_ML_output}       
\end{figure}

A machine learning model, based on the BDT and implemented with the XGBoost algorithm, is employed to further enhance the J/$\psi$ signal significance. Half of the signal and background samples are used for model training, while the other half is reserved for testing. Once the training is complete, the resulting model is applied to both the training and testing samples. For each electron–positron pair, the model assigns an output value between 0 and 1 based on its input features. Values closer to 1 indicate a higher probability of being signal, while values closer to 0 suggest a higher likelihood of being background. The resulting distributions of model outputs are shown in Fig.~\ref{fig_ML_output}. The red band and red solid circles represent distributions for the training and testing signal samples, respectively, while the blue band and blue solid circles correspond to the training and testing background samples. As expected, the model outputs for the signal samples are closer to 1, while those for the background samples are closer to 0. Furthermore, the consistency between the training and testing samples demonstrates that no overfitting has occurred.

\begin{figure}[htbp]
\centering
\includegraphics[width=0.46\textwidth,clip]{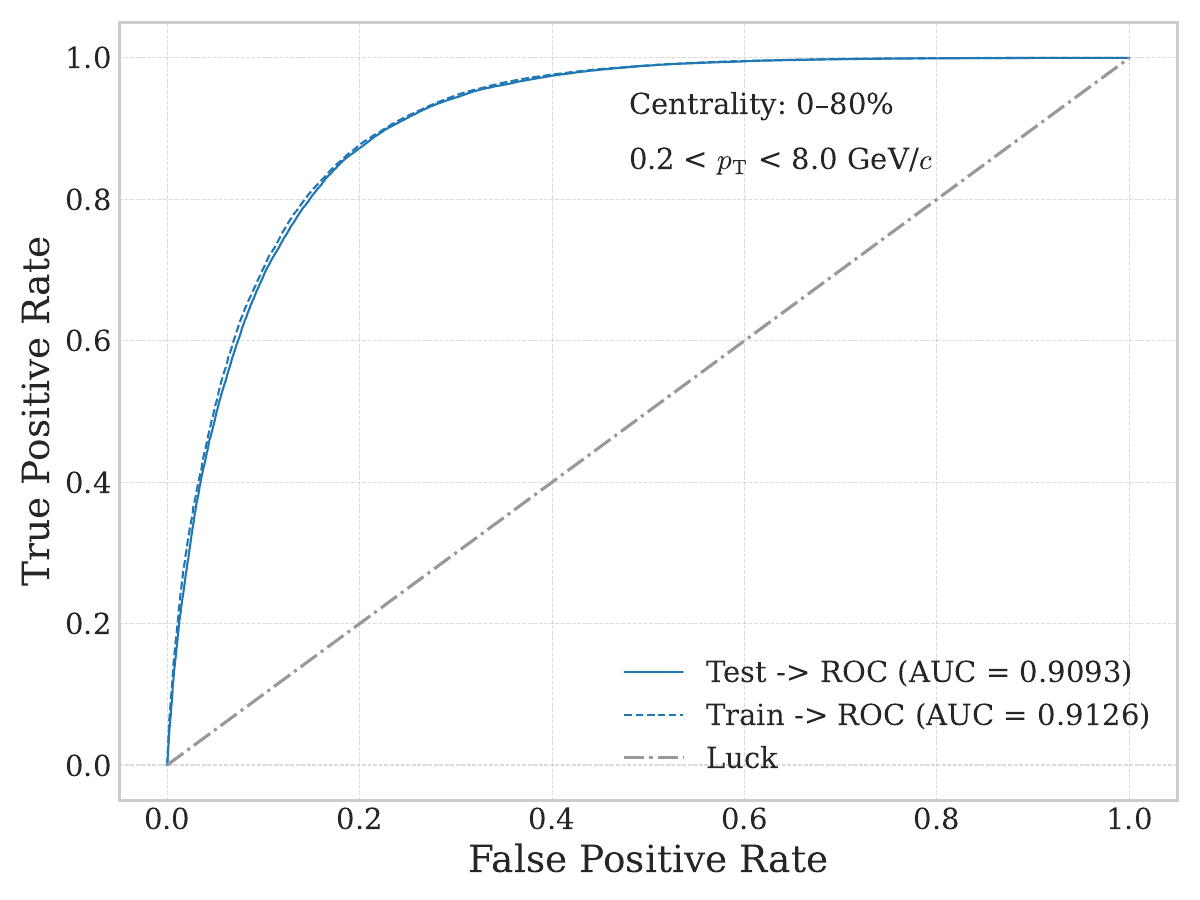}
\caption{ROC curves for the training (dashed) and testing (solid) samples.}
\label{fig_ML_ROC}       
\end{figure}

In real-life applications, it is common to select a specific BDT criterion or cut, where events with model output values above the criterion are classified as signal, and those below are considered background. The signal efficiency at a given criterion is defined as the ratio of the number of signal events above the criterion to the total number of signal events. Taking this efficiency as the $y$-axis value and the corresponding background efficiency as the $x$-axis value yields a coordinate point ($x$, $y$). As the criterion is varied from 0 to 1, these points trace out a curve known as the Receiver Operating Characteristic (ROC) curve. The $x$ and $y$ values correspond to the False Positive Rate (FPR) and True Positive Rate (TPR), respectively. The ROC curves for both the training and testing samples are shown in Fig.~\ref{fig_ML_ROC}. Additionally, the Area Under the Curve (AUC) provides a quantitative measure of the model’s classification performance. The closer the AUC to 1, the better the performance. In this work, the AUC reaches 0.91, indicating strong discriminative power. The nearly identical AUC values for the training and testing samples further confirm quantitatively that the model is not overfit. 

\section{Model Validation via Self-Consistency Checks}
\label{sect::modelvalidation}

In addition to ensure that the trained model does not suffer from overfitting, as demonstrated in the previous section, one also need to verify that it can reproduce the BDT cut efficiencies for both signal and background in real data. This can be done through two self-consistency checks. 

\begin{figure}[htbp]
\centering
\includegraphics[width=0.46\textwidth,clip]{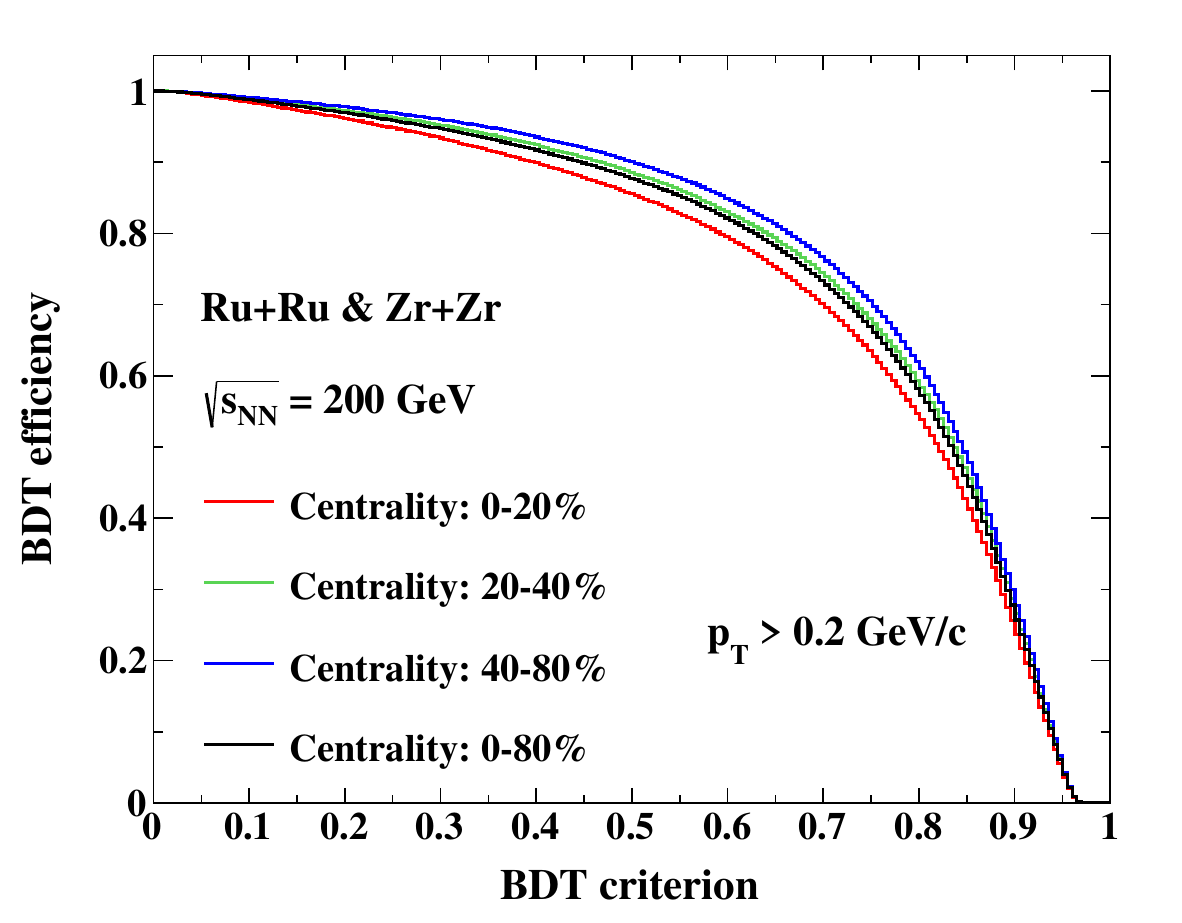}
\caption{BDT cut efficiencies for J/$\psi$ with $p_{\rm T}>0.2$ GeV/$c$ in different centrality classes of 200 GeV Ru+Ru and Zr+Zr collisions.}
\label{fig_ML_eff}       
\end{figure}

\begin{figure}[htbp]
\centering
\includegraphics[width=0.46\textwidth,clip]{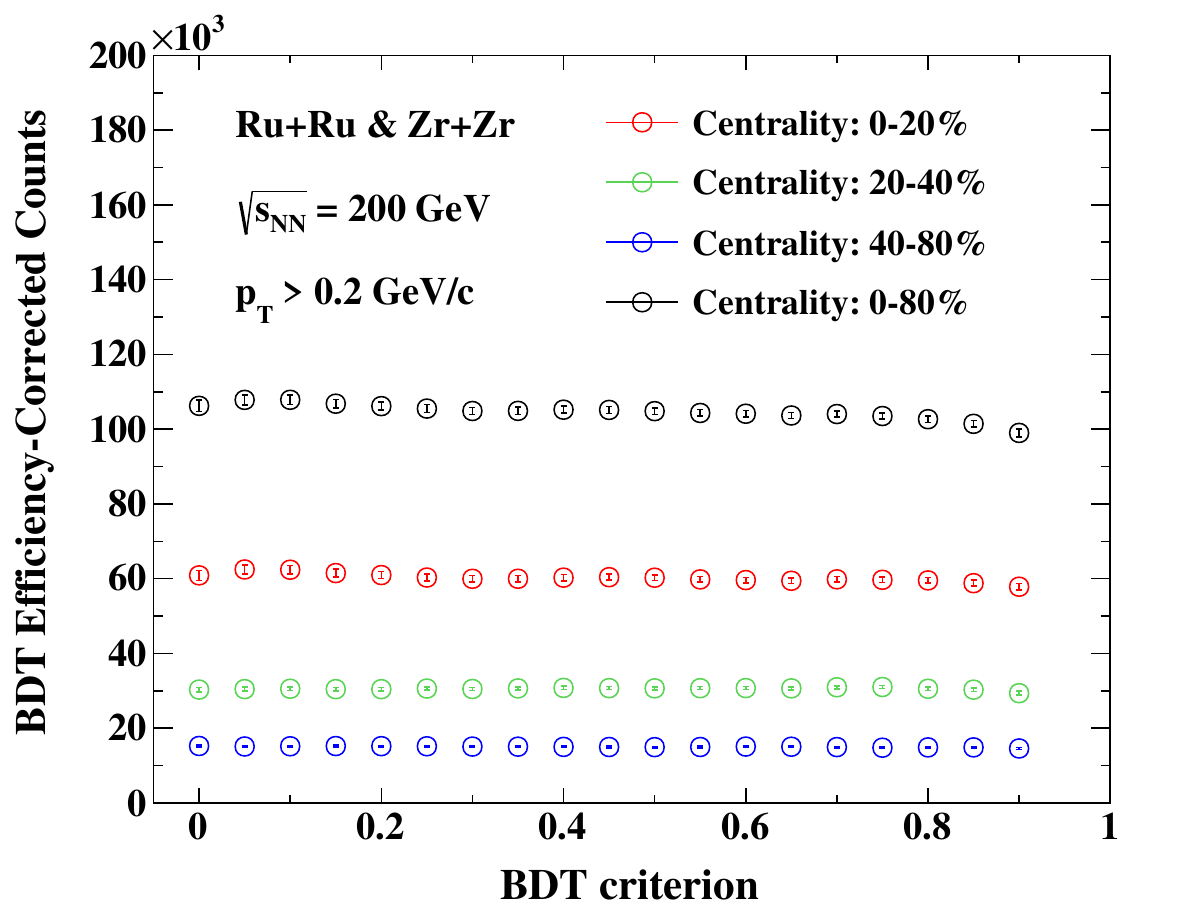}
\caption{J/$\psi$ counts, corrected by BDT cut efficiencies, as a function of BDT criterion in different centrality classes of 200 GeV Ru+Ru and Zr+Zr collisions.}
\label{fig_ML_counts}       
\end{figure}

First, we check the J/$\psi$ count, corrected for the BDT cut efficiency, against the BDT criterion, which is expected to remain constant as experimental results should be independent of specific cuts used. The BDT cut efficiency for J/$\psi$ is obtained from the signal's model output distribution (Fig. \ref{fig_ML_output}) and shown in Fig.~\ref{fig_ML_eff} as a function of the BDT criterion for different centralities of Ru+Ru and Zr+Zr collisions. Here, collision centrality describes the degree of overlap between the two incoming nuclei: 0-20\% corresponds to collisions with largest overlap and constituting 20\% of hadronic cross section, while 40-80\% represents those of smaller overlap. After applying the preselection cuts (see Tab.~\ref{table_pre_pid}) and different BDT cuts, the raw J/$\psi$ counts can be extracted from real data using the reconstruction procedure introduced in Sec. \ref{sect::jpsireco}. Then, Fig.~\ref{fig_ML_counts} shows the J/$\psi$ counts, corrected for the BDT cut efficiency, as a function of the BDT criterion in four different centrality classes. Although the BDT cut efficiency changes by an order of magnitude as the BDT criterion varies from 0 to 0.9, the efficiency-corrected signal counts remain mostly constant against BDT criterion, with less than 2\% variations. This indicates that the  BDT cut efficiency for J/$\psi$ as obtained from the machine learning model precisely represents that in real data. This is only achievable if the signal in real data shares the same model output distribution as the simulated signal sample used for training, confirming that the corrections applied to the various features of the signal training sample, as described in Sec. \ref{sect::trainingsample}, work well.

\begin{figure}[htbp]
\centering
\includegraphics[width=0.46\textwidth,clip]{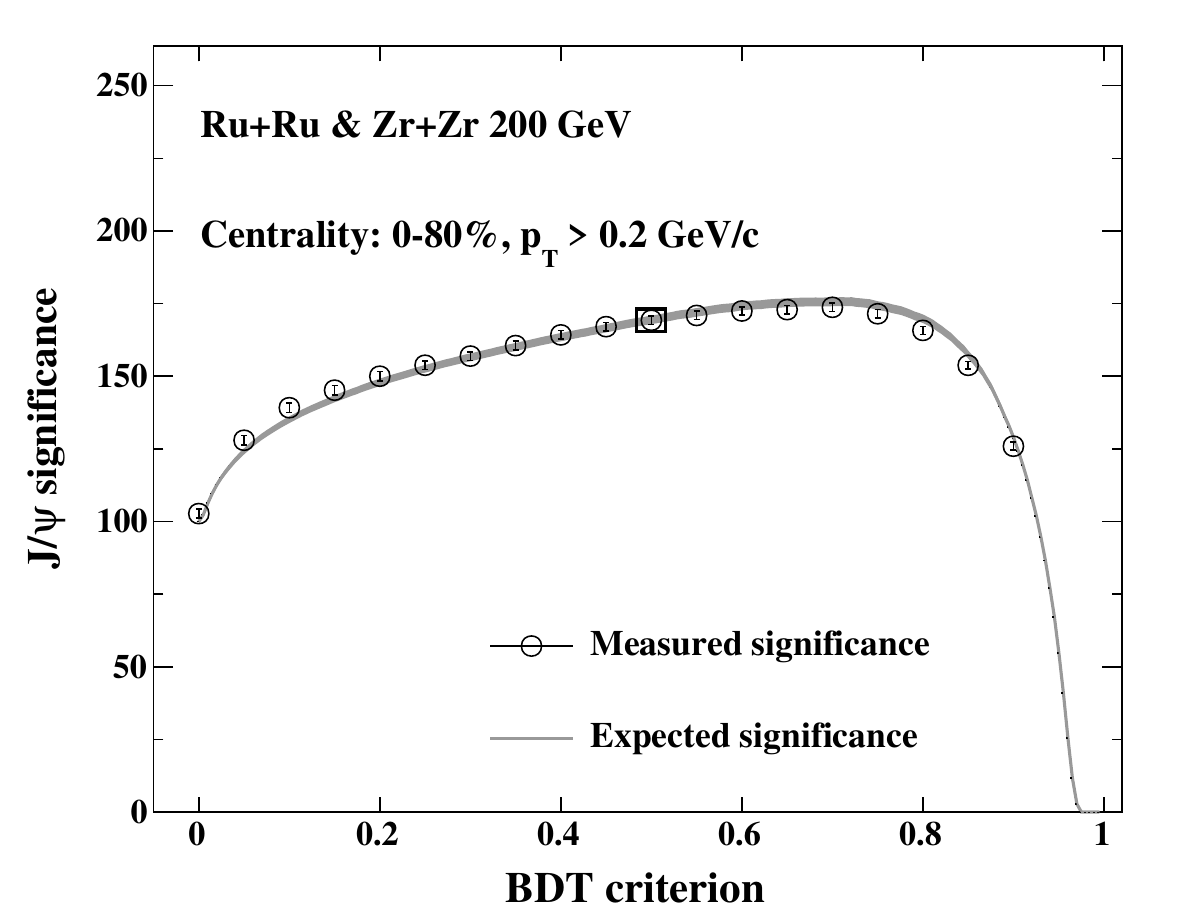}
\caption{J/$\psi$ significance as a function of BDT criterion. The black open circles represent the J/$\psi$ significance directly obtained from real data while the black band shows the expected significance estimated from the machine learning model. The two distributions are aligned at BDT value of 0.5 as indicated by the black open square.}
\label{fig_ML_workpoint}       
\end{figure}

A more stringent test of the model performance can be done through a comparison of the J/$\psi$ significance as a function of the BDT criterion between the machine learning model and real data as such a comparison involves efficiencies for both signal and background. 
The expected J/$\psi$ signal significance from the machine learning model is calculated as follows:
\begin{equation}
\setlength{\abovedisplayskip}{6pt}
\setlength{\belowdisplayskip}{6pt}
\begin{gathered}
    S(x) = S_{0} \cdot \text{eff}_{s}(x) / \text{eff}_{s}(x_{0}),\\
    B(x) = B_{0} \cdot \text{eff}_{b}(x) / \text{eff}_{b}(x_{0}),\\
    \text{Significance}_{\text{expected}}(x) = S(x) / \sqrt{S(x) + B(x)}.\\
\end{gathered}
\label{func-workpoint}
\end{equation}
Here, $x$ represents the BDT criterion, and $\text{eff}_{s}(x)$ and $\text{eff}_{b}(x)$ correspond to the signal and background BDT criterion efficiencies, respectively. $S(x)$ and $B(x)$ denote the signal and background counts, and $\text{Significance}_\text{expected}(x)$ denotes the J/$\psi$ significance. $S_{0}$ and $B_{0}$ represent the J/$\psi$ signal and background counts at a specific BDT criterion of $x_0$. The choice of this BDT criterion is not unique and serves only as a reference point for aligning with real data. The resulting J/$\psi$ significance as a function of the BDT criterion is shown in Fig.~\ref{fig_ML_workpoint} as the gray band. It rises and falls as the BDT criterion increases from 0 to 0.9, resulting from the interplay of the BDT cut efficiencies for signal and background. The distribution for the machine learning model is compared to that extracted from real data shown as open circles. The two distributions are aligned at the BDT criterion of 0.5, {\it i.e.}, $x_{0}=0.5$ in Eq. \ref{func-workpoint}, to facilitate the comparison. The consistency between the expected significance from trained model and the measured significance in real data demonstrates that both the signal and background samples used for training accurately represent their behaviors in real data.

The trend of the expected significance as a function of the BDT criterion from the trained model can also be utilized to determine the optimal BDT criterion. Compared to directly scanning the real data to select the optimal criterion, using simulations avoids influences of fluctuations in real data, which may introduce biases in yield measurements, especially for low-statistics signals. As shown in Fig. \ref{fig_ML_workpoint}, the J$/\psi$ significance reaches a plateau around the BDT criterion of 0.7, which is chosen as the optimal criterion. 
Figure~\ref{fig_invariant} (b) shows the raw signal extraction with the optimal BDT criterion, where the signal significance reaches 174, exhibiting a 69\% improvement compared to using only preselection cuts (Fig.~\ref{fig_invariant} (a)) and a 25\% enhancement over the result obtained with straight cuts (see Tab. \ref{table_pid} in the Appendix) as shown in Fig.~\ref{fig_invariant} (c).

The self-consistency checks, based on the dependence of the efficiency-corrected signal count and signal significance on the BDT criterion, can also be used for scrutinizing features. If the inclusion of a specific feature in the model training disrupts the consistency between trained model and real data, such a feature would need to be either corrected or excluded. For example, Fig.~\ref{fig_ML_counts_compare} in the Appendix shows the results without applying the CDF mapping correction to the $E_{0}/p$ feature. In this case, the corrected signal counts exhibit increasingly large deviations from the expected flat distribution as the BDT criterion increases, reaching 29\% at the BDT criterion of 0.9. Similarly, the expected J/$\psi$ signal significance shows a clear inconsistency to that obtained from real data, as displayed in Fig. \ref{fig_ML_workpoint_noCor} in the Appendix. Such discrepancies clearly demonstrate the necessity and effectiveness of the feature correction procedures introduced in Sec. \ref{sect::trainingsample}.

\section{Summary}
\label{sect::summary}

In this study, we develop two methods, CDF mapping and shift-and-scale, to align feature distributions in simulations with those in real data while preserving these correlations. These approaches address a common challenge in applying machine learning to experimental analyses: discrepancies between simulated and real data that can introduce systematic biases if uncorrected. Using the measurement of J/$\psi$ yield in 200 GeV Ru+Ru and Zr+Zr collisions with a BDT algorithm as an example, we demonstrate the validity and effectiveness of these corrections. Self-consistency checks, based on comparisons of efficiency-corrected J/$\psi$ yields and J/$\psi$ signal significance between trained model and real data, confirm their necessity. These methods are broadly applicable, enhancing the reliability of machine learning in high-energy particle and nuclear physics.


\section*{Acknowledgments}


We would like to express our gratitude to the STAR Collaboration and the RHIC Operations Group for providing the valuable data utilized in this work. Additionally, we sincerely thank the RCF facility at Brookhaven National Laboratory for their support in offering the computational resources essential for this study. We also thank Pengzhong Lu and Zhenjun Xiong for their helpful discussions and assistance.

\appendix
\section{Results without the $E_{0}/p$ correction}

\begin{figure}[htbp]
\centering
\includegraphics[width=0.46\textwidth,clip]{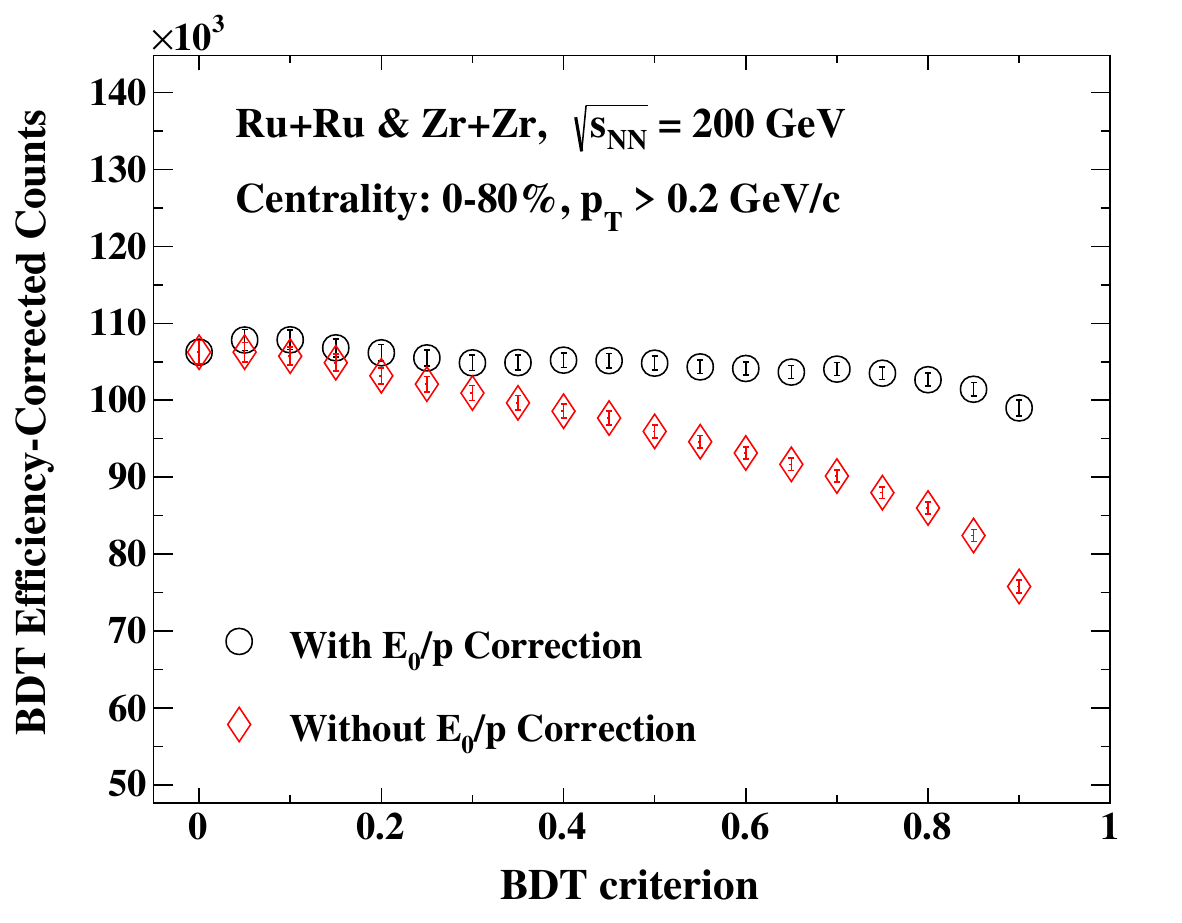}
\caption{J/$\psi$ counts, corrected for the BDT cut efficiencies, as a function of BDT criterion, for scenarios that the signal sample is incorporated with (circles) or without (diamonds) the $E_{0}/p$ correction.}
\label{fig_ML_counts_compare}      
\end{figure}

Figure \ref{fig_ML_counts_compare} shows the BDT efficiency-corrected J/$\psi$ yield as a function of the BDT criterion for the two cases that the $E_{0}/p$ feature distribution in the training sample is aligned or not aligned with that in real data.

\begin{figure}[htbp]
\centering
\includegraphics[width=0.46\textwidth,clip]{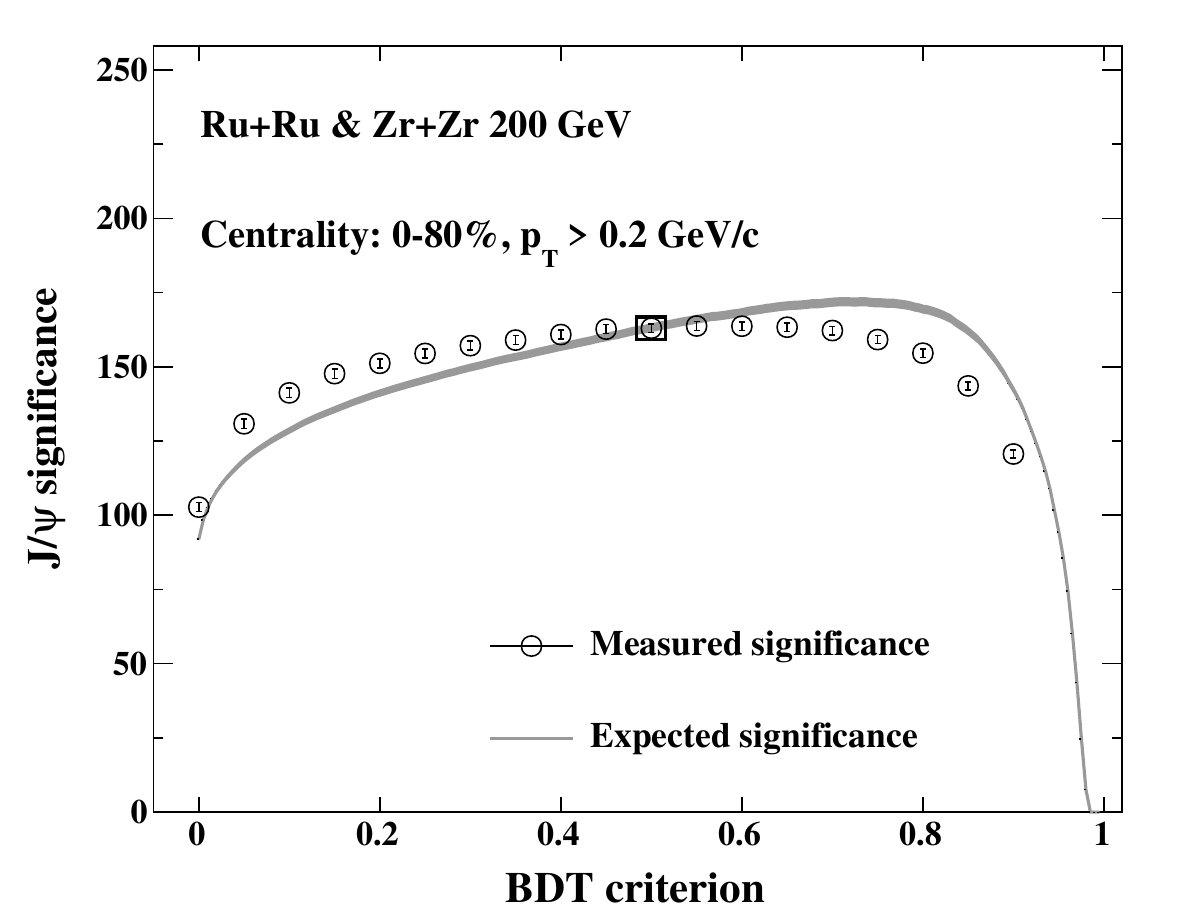}
\caption{Expected J/$\psi$ significance as a function of the BDT criterion from the machine learning model trained using the signal sample without the $E_{0}/p$ correction. It is compared to the measured significance in real data (open circles). The two distributions are aligned at the BDT criterion of 0.5, indicated by the open square, to facilitate the comparison.}
\label{fig_ML_workpoint_noCor}       
\end{figure}

Figure \ref{fig_ML_workpoint_noCor} shows the comparison of the expected J/$\psi$ signal significance from trained model without applying the CDF mapping to the $E_{0}/p$ distribution to the measured significance in real data.

\section{Straight electron identification cuts}
\begin{table*}[htbp]
    \centering
    \caption{Straight cuts used for electron identification.}
    \begin{tabular*}{\textwidth}{@{\extracolsep{\fill}} ccc}
        \hline
        Track $p_{\rm T}$ & Detectors used & Electron identification cuts \\
        \hline
        $p_{\rm T} \leq 1.0$ GeV/$c$ & TPC, TOF & {\footnotesize\makecell{$\left|1/\beta - 1 \right|<0.025$; \\ for $p > 0.8$ GeV$/c$: -0.75 < n$\sigma_{e}$ < 2; \\ for $p \leq 0.8 $ GeV$/c$: (3$\times$ p - 3.15) < n$\sigma_{e}$ < 2}} \\
        \hline
        \multirow{7}{*}{$p_{\mathrm{T}} > 1.0$ GeV/$c$} & \makecell{TPC, TOF and \\ BEMC no matching} & {\footnotesize\makecell{$\left|1/\beta - 1 \right|<0.025$; \\  $-0.75 < n\sigma_{e} < 2$ }} \\
          & \multirow{3}{*}{\makecell{TPC, BEMC and \\ TOF no matching}} & \multirow{3}{*}{\footnotesize\makecell{$-1 < n\sigma_{e} < 2$; \\ $0.5 < E_{0}/p < 1.5$}} \\
          & & \\
          & & \\
         & \makecell{TPC, TOF and \\ BEMC} & {\footnotesize\makecell{$\left|1/\beta - 1 \right|<0.025$; \\  $-1.5 < n\sigma_{e} < 2$; \\$0.5 < E_{0}/p < 1.5$ }} \\
        \hline
    \end{tabular*}
    \label{table_pid}
\end{table*}

\bibliography{cite2.bib}

\end{document}